# Critical Properties of the Models of Small Magnetic Particles of the Antiferromagnet MnF$_2$

V. A. Mutailamov [a,*], A. K. Murtazaev [a,b], and M. A. Magomedov [a,b]

[a] *Institute of Physics, Dagestan Scientific Center, Russian Academy of Sciences, ul. Yaragskogo 94, Makhachkala, 367003 Russia*

[b] *Dagestan State University, ul. M. Gadzhieva 43a, Makhachkala, 367025 Russia*

*e-mail: vadim.mut@mail.ru

ABSTRACT

The static critical behavior of the models of small magnetic particles of the real two-sublattice antiferromagnet MnF$_2$ is investigated by the Monte Carlo method taking into account the interaction of the second nearest neighbors. Systems with open boundaries are considered to estimate the influence of the sizes of particles on the pattern of their critical behavior. The behavior of thermodynamic functions in the phase transition region is investigated. The data on the maxima of the temperature dependences of heat capacity and magnetic susceptibility are shown to be insufficient to unambiguously determine the effective temperture of the phase transition in the models of small magnetic particles. This requires an additional investigation of the spatial orientation of the sublattice (sublattices) magnetization vector for the models under study.

## 1. INTRODUCTION

In modern condensed matter physics, special attention is given to studying the phase transitions and critical properties of small particles of magnetic materials containing from hundreds to tens of thousands of atoms (spins, ions). Such systems possess a number of interesting properties that macroscopic objects do not possess. Studying the influence of finite sizes of small particles on particular physical properties of a material in the phase transition region is of great interest.

Greatly simplified models are generally considered when small particles are studied theoretically, although they are fairly diverse. In many cases, despite the model simplicity, the analytical theories encounter great difficulties. In addition, the passage to the thermodynamic limit is impossible for such systems. An experimental study of the properties of small particles also runs into considerable difficulties: the number of interacting elements, the particle shape, the presence of impurities and defects, the appearance of an oxide shell, and the effects attributable to interparticle interactions often cannot be controlled directly. The numerical simulation methods are devoid of these shortcomings. Developing the models of small magnetic structures and the algorithms for their investigation is an important task of modern computational physics. The results obtained by the methods of computational physics are not only competitive in accuracy with other theoretical methods but often surpass them [1, 2].

Until now, the systems that can be described by simple classical models of the first approximation, such as the Ising model, the Heisenberg model, the *XY* model, etc., have generally been the main objects of investigation. In real materials, however, apart from the exchange interaction, there are also various factors disregarded by the classical models of the first approximation. Uniaxial anisotropy, dipole–dipole interaction, the presence of a free surface, allowance for the interaction between particles of the second coordination sphere can exert a significant influence on the pattern of static and dynamic critical behavior of both macroscopic and small-sized magnets [3–5].

## 2. THE MODEL AND THE METHOD

In this paper, we investigated the static critical behavior of the models of small particles of the real two-sublattice antiferromagnet $MnF_2$ by taking into account the interaction of the second nearest neighbors. When constructing the model, we took into account the main magnetic and crystallographic peculiarities of the real material [6, 7]. The manganese atoms in $MnF_2$ form a body-centered tetragonal structure with the lattice constants a = 4.87Å and $c$ = 3.31Å. The main exchange interactions here are the antiferromagnetic one along the [111] direction with the constant $J_1/k_B = -1.76K$ and the weak ferromagnetic one along the [001] direction with the constant $J_2/k_B = 0.3 \pm 0.1K$. The exchange interaction along the [100] and [010] directions is close to zero. Thus, the ferromagnetic interaction between the nearest spins along the c axis turns out to be much weaker than the antiferromagnetic interaction of the central spin with the spins in the lattice corners. The spin is $S$ = 5/2 [8], which apparently allows the semiclassical ($S = \infty$) Heisenberg model to be used to describe this system. The $MnF_2$ lattice structure is schematically shown in Fig. 1.

In real MnF2 samples, there is a strong uniaxial anisotropy directed along the *c* axis. As the anisotropy constant, we used $D_A = 0.0591|J_1|$ corresponding to macroscopic $MnF_2$ samples. We have failed to find how this quantity is modified for small particles in the literature. According to the experimental data, the phase transition from the antiferromagnetic state to the paramagnetic one is observed in $MnF_2$ samples at the Néel temperature $T_N$ = 67.34 K [7, 10].

Thus, given the peculiarities of the real material, the $MnF_2$ model Hamiltonian can be represented as

$$H = -\frac{1}{2}J_1\sum_{i,j}(\vec{S}_i\vec{S}_j) - \frac{1}{2}J_2\sum_{i,k}(\vec{S}_i\vec{S}_k) - D_A\sum_i(S_i^z)^2, \quad |\vec{S}_i| = 1. \quad (1)$$

where the first sum allows for the exchange interaction of the Mn ions with the ions located in the corners of a unit cell ($J_1$ < 0), the second sum allows for the exchange interaction between the neighbors along the c axis ($J_2$ > 0, |$J_2/J_1$| = 0.170), and the third sum allows for the uniaxial anisotropy.

During our studies, we simulated particles with a tetragonal shape containing $L \times L \times L$ unit cells in each crystallographic direction. The particles were oriented in space in such a way that the *z* coordinate axis coincided with the crystallographic *c* axis and the *x* axis coincided with the crystallographic *a* axis. To estimate the influence of the sizes of particles on the pattern of their critical behavior, we considered systems with open boundaries. We performed our simulations by the Monte Carlo method based on the standard Metropolis algorithm [1, 2, 11].

The parameters of the investigated particles are given in Table 1. For all linear sizes *L*, the table provides the total number of spins *N*, the number of surface spins $N_S$, and the fraction of surface spins $P_S$. The spins without a complete set of first- or second-order nearest neighbors were taken as the surface ones.

During our simulations, we discarded the initial nonequilibrium segment of the Markov chain, which is definitely longer that the relaxation time of the particle being investigated, to bring the spin system into a state of thermodynamic equilibrium. The length of the nonequilibrium segment was chosen depending on the temperature and for all linear sizes was 100 000 Monte Carlo steps per spin far from the critical region and 200 000 steps in the phase transition region. The ensemble-averaged thermodynamic quantities were calculated in the equilibrium state. The length of the equilibrium segment for all linear sizes was from 500 000 to 1 000 000 Monte Carlo steps per spin depending on the temperature region. In addition, to improve the statistics at each temperature for all linear sizes, we performed up to five simulations for various initial spin configurations whose results were then averaged between themselves. As the order parameter of the model of the antiferromagnet MnF$_2$, we used the magnitude of the antiferromagnetism vector $M$ calculated as the magnetization difference between its two sublattices:

$$M = \sqrt{(M_{1x} - M_{2x})^2 + (M_{1y} - M_{2y})^2 + (M_{1z} - M_{2z})^2}, \quad (2)$$

where Mnk is the kth projection of the magnetization of the nth sublattice. To monitor the temperature dependence of the behavior of heat capacity and magnetic susceptibility, we used the fluctuation relations [12]

$$C = (NK^2)(\langle U^2 \rangle - \langle U \rangle^2), \quad (3)$$

$$\chi = (NK)(\langle M^2 \rangle - \langle M \rangle^2), \quad (4)$$

where $K = |J_1|/k_B T$, $N$ is the number of particles, and $U$ is the internal energy.

The temperature dependences of the thermodynamic parameters were calculated with the temperature step $\Delta T = 0.1 \, k_B T/|J_1|$ far from the phase transition region and with the step $\Delta T = 0.02 \, k_B T/|J_1|$ near the critical point.

When converting the temperature from dimension less units $k_B T/|J_1|$ to kelvins, the fact that we used the classical Hamiltonian of the Heisenberg model (1) during our simulations should be taken into account. The behavior of the quantum Heisenberg model with a large spin ($S \geq 2$) is similar to the behavior of the classical Heisenberg model if the vectors equal in magnitude to are used as the spins [13]. In this case, the exchange interactions do not change. Since we used vectors of unit length in Eq. (1), the exchange interaction should be renormalized to $\sqrt{S(S+1)}$ to recalculate the temperature. Thus, given the form of Hamiltonian (1), we calculated the temperature $T$ in kelvins for MnF$_2$ from the formula

$$T = t \cdot 2 S(S+1) |J_1|/k_B, \quad (5)$$

where t is the temperature in arbitrary units of the exchange integral that we used during our simulations. For clarity, we provide the temperature both in dimensionless units $k_B T/|J_1|$ and in kelvins on all of the graphs presented below.

## 3. RESULTS

Figure 2 presents our temperature dependences of the order parameter for particles with various linear sizes. The vertical dotted line on the graph marks the phase transition temperature $T_N = 67.34 K$ for real MnF$_2$ samples. As can be seen from the figure, there are no size effects at a temperature close to zero. As the temperature rises, a difference in the curves obtained for different linear sizes appears. This behavior of the order parameter in the low-temperature region is typical of the models of small magnetic particles with open boundaries [14]. In the high-temperature region, the size effects manifest themselves as the "tails" traditional for the Monte Carlo method. As the linear sizes increase, the tails are reduced and the order parameter approaches zero.

The temperature dependence of heat capacity for particles with various linear sizes is presented in Fig. 3. As can be seen from the figure, the heat capacity maxima near the critical region have smoothed peaks instead of the characteristic features. The smoothing decreases with increasing linear particle sizes, while the absolute value of the maxima, accordingly, increases. This peculiarity in the behavior of some derivatives of the free energy is typical of small particles and stems from the fact that the correlation length near the critical region is equal to the linear particle sizes [11].

The heat capacity maxima are shifted in temperature relative to the phase transition temperature $T_N$ of the real material (an infinite system) toward low temperatures. The shift increases with decreasing linear sizes. The shift in critical temperature stems from the fact that the fraction of surface spins without a complete set of nearest neighbors is large in small particles. As a result, disorder arises at lower temperatures [11].

Figure 4 presents the temperature dependence of magnetic susceptibility for particles with various linear sizes. Here, a smoothing of the peaks near $T_N$ and their shift in temperature are also observed. Note that the temperatures at which the heat capacity and magnetic susceptibility maxima occur do not coincide between themselves for identical linear sizes. The magnetic susceptibility maxima occur at higher temperatures. This noncoincidence is typical of both the models of macroscopic samples and the models of small particles [14]. The derived temperatures of the heat capacity and magnetic susceptibility maxima for all linear sizes are given in Table 2. Note that the difference in temperature at the heat capacity and magnetic susceptibility maxima decreases with increasing particle sizes. The noncoincidence of the maxima makes the choice of an effective phase transition temperature in the models of small particles ambiguous. We performed the following procedure to obtain a rigorous criterion.

During our simulations, we calculated the angle θ between the direction of the magnetization vector for each sublattice and the $z$ axis at each Monte Carlo step per spin. The data obtained were then averaged over the entire length of the Markov chain. Figure 5 presents the temperature dependence of the angle θ obtained in this way for particles with the linear sizes $L = 14$ and 22. Note that the lattice numbering is arbitrary. We chose the sublattices with smaller and larger angles θ as the first and second ones, respectively. As can be seen on the graph, the magnetic moments of the sublattices at low temperatures are oriented along the $z$ axis in mutually opposite directions. At some temperature, the order disappears completely and the particle passes from the antiferromagnetic state into the paramagnetic one. As follows from Table 2, the ordering temperature occurs at the maximum of the temperature dependence of heat capacity for a given particle. Thus, it is apparently desirable to determine the effective phase transition temperature in the models of small particles from the position of the maximum of the temperature dependence of heat capacity. It can also be seen from the figure that disorder occurs in a fairly narrow temperature range that decreases with increasing linear sizes of the system.

According to the theory of finitesize scaling [1, 2], the effective phase transition temperature $T_N(L)$ for a particle with finite linear sizes $L$ differs from the phase transition temperature $T_N$ for an infinite system by [1]

$$T_N(L) = T_N - \lambda L^{-1/\nu}, \qquad (6)$$

where λ is some constant and ν is the static critical index of the correlation length. The derived dependence of the phase transition temperature $T_N(L)$ on linear sizes $L$ is presented in Fig. 6. The temperature at which the heat capacity maximum occurs was used as $T_N(L)$.

Linear fitting of this dependence by Eq. (6) allows the critical temperature of an infinite system to be calculated from the intersection of the interpolation straight line with the temperature axis at the point $L^{-1/\nu} \to 0$. The linear fit was made by the least-squares method and is indicated in the figure by the solid line. Investigation of the model of macroscopic $MnF_2$ samples with allowance made for the interaction of the first nearest neighbors showed that the presence of a strong uniaxial anisotropy leads to the Ising universality class of the static critical behavior of this model [15]. Therefore, in our calculations based on Eq. (6), we used ν = 0.6298(8) typical of the classical three-dimensional Ising model [1] as the static critical index of the correlation length. As a result, we obtained the critical temperature $T_N = 2.20(2)\, k_B T/|J_1|$ for an infinite system. Note that this value in kelvins corresponds to $T_N = 67.8(6)$ K, which is close to the phase transition temperature for real $MnF_2$ samples.

## 4. CONCLUSIONS

Our studies show the efficiency of applying the methods of computational physics for studying the static critical behavior of the models of small particles of real magnetic materials. The Monte Carlo method allows the influence of size effects on the thermophysical properties of the models of small particles to be investigated. Since the maxima of the temperature dependences of heat capacity and magnetic susceptibility for identical linear sizes occur at different temperatures, these data are insufficient to unambiguously determine the effective temperature of the phase transition for the models of small magnetic particles. An additional investigation of the spatial orientation of the sublattice (sublattices) magnetization vector for the models under study is needed to determine the ordering temperature more rigorously.

## ACKNOWLEDGMENTS

This study was financially supported by the Ministry of Education and Science of the Russia Federation (contract no. 14.B37.21.1092) and the Russian Foundation for Basic Research (project nos. 13-02-00220_A_2013 and 12-02-96504_r_yug_a).

**Table 1**. Parameters of the investigated particles

| L  | N      | $N_S$ | $P_S$, % |
|----|--------|-------|----------|
| 10 | 1 729  | 650   | 37.59    |
| 14 | 4 941  | 1 354 | 27.40    |
| 18 | 10 745 | 2 314 | 21.54    |
| 22 | 19 909 | 3 530 | 17.73    |

**Table 2.** Temperatures of the heat capacity $T_{max}^{C}$ and magnetic susceptibility $T_{max}^{\chi}$ maxima for various linear sizes

| L  | $T_{max}^{C}$, $k_B T/|J_1|$ | $T_{max}^{\chi}$, $k_B T/|J_1|$ | $T_{max}^{C}$, K | $T_{max}^{\chi}$, K |
|----|------------------------------|--------------------------------|------------------|---------------------|
| 10 | 1.84(2)                      | 1.96(2)                        | 56.7(6)          | 60.4(6)             |
| 14 | 1.98(2)                      | 2.04(2)                        | 61.0(6)          | 62.8(6)             |
| 18 | 2.06(2)                      | 2.10(2)                        | 63.4(6)          | 64.7(6)             |
| 22 | 2.10(2)                      | 2.12(2)                        | 64.7(6)          | 65.3(6)             |

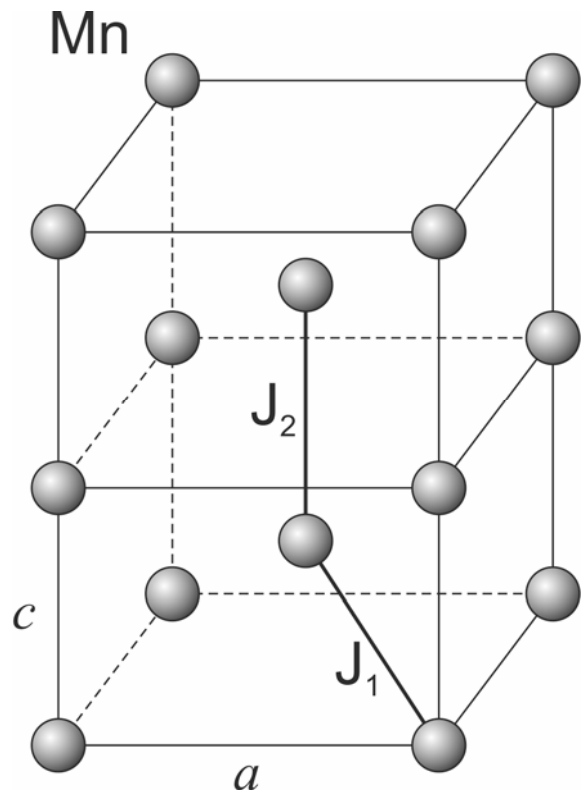

**Fig. 1**. Schematic view of the magnetic Mn ions in the MnF$_2$ lattice.

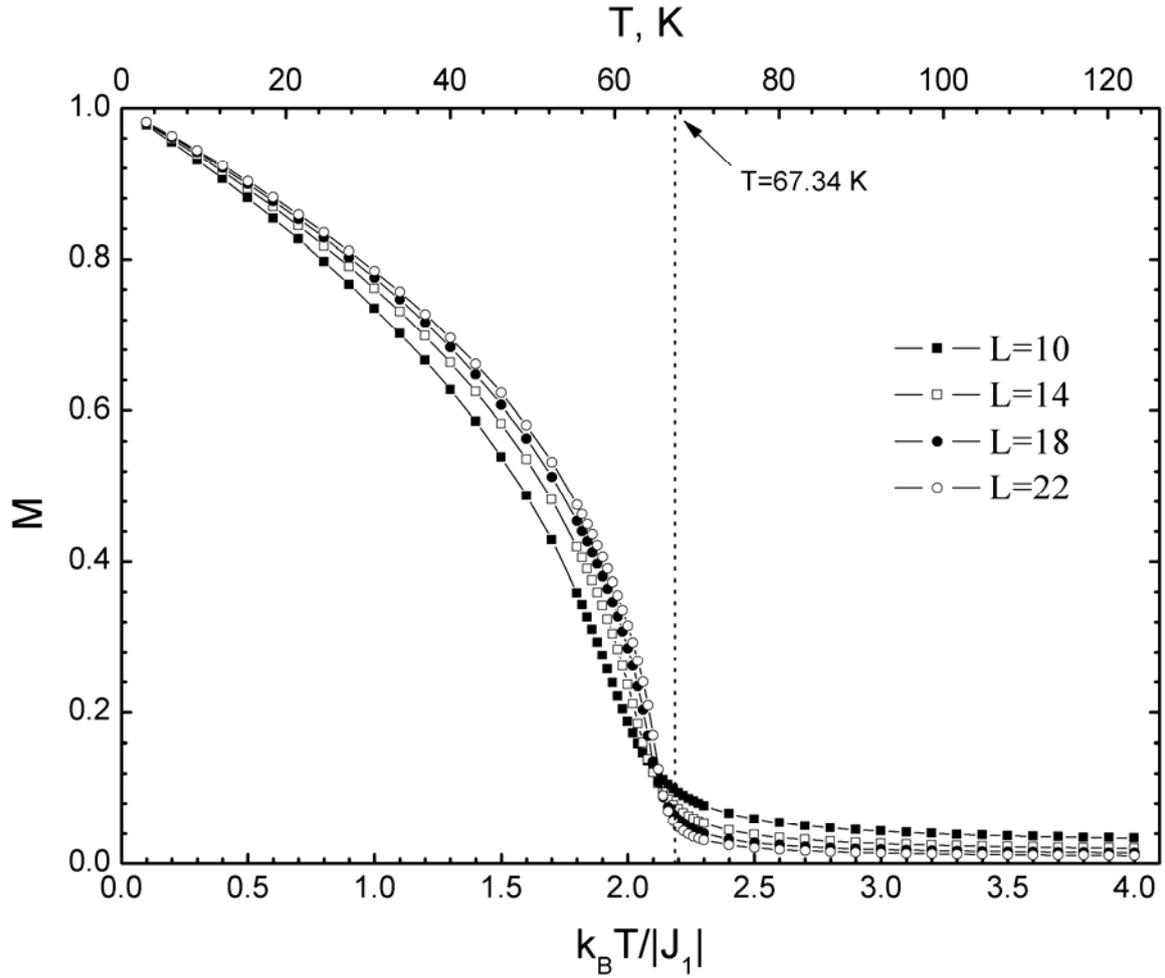

**Fig. 2**. Temperature dependence of the magnitude of the antiferromagnetism vector *M* (order parameter) for particles with various linear sizes *L*.

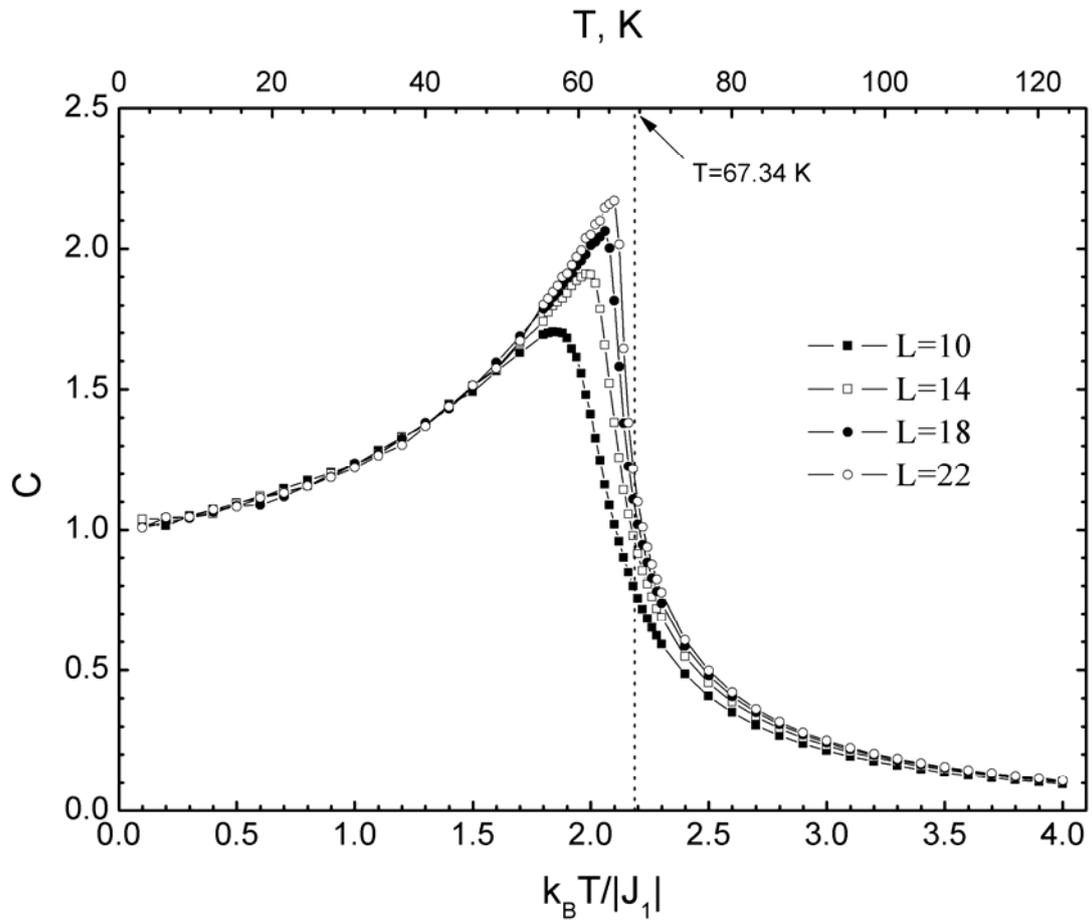

**Fig. 3.** Temperature dependence of heat capacity *C* for particles with various linear sizes *L*.

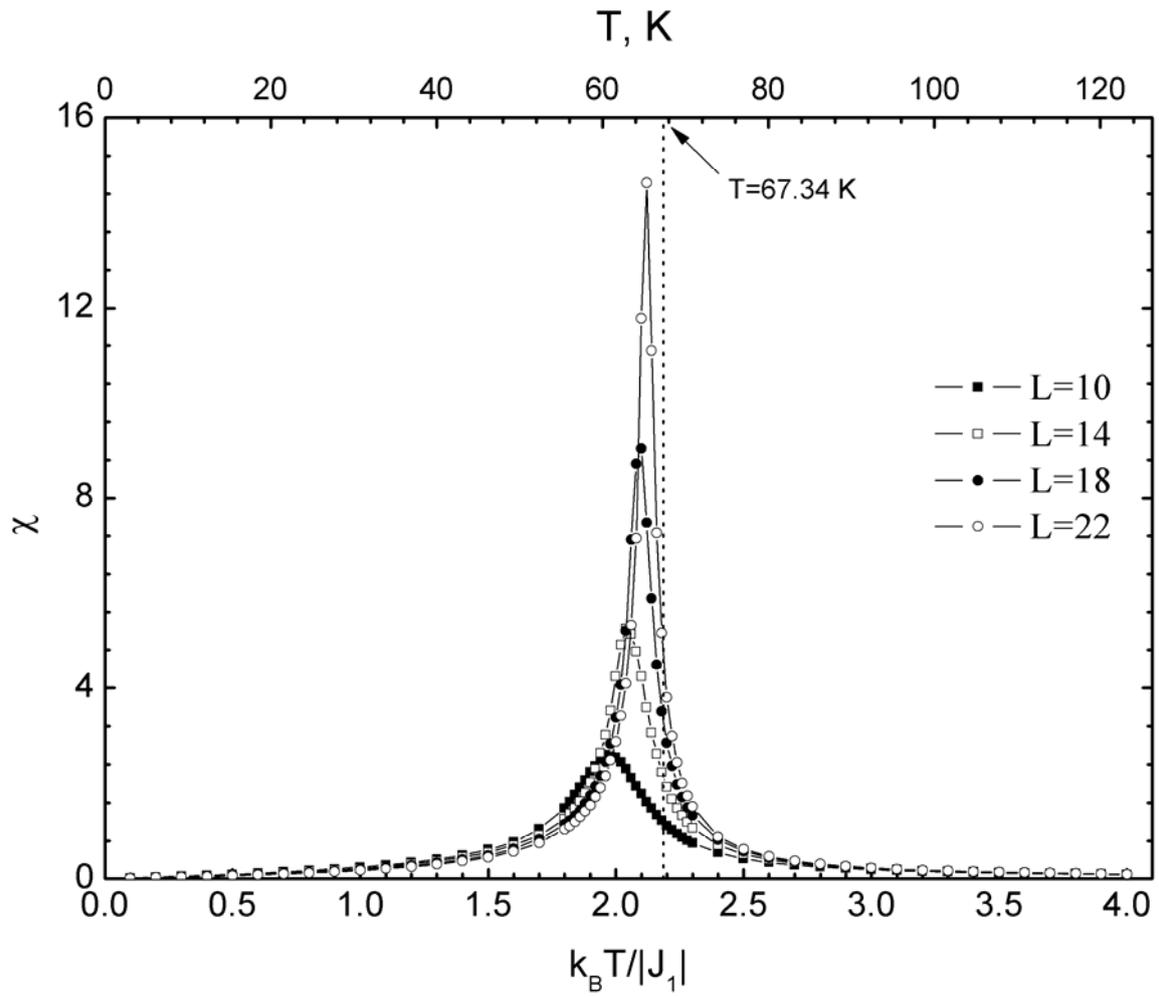

**Fig. 4.** Temperature dependence of magnetic susceptibility χ for particles with various linear sizes *L*.

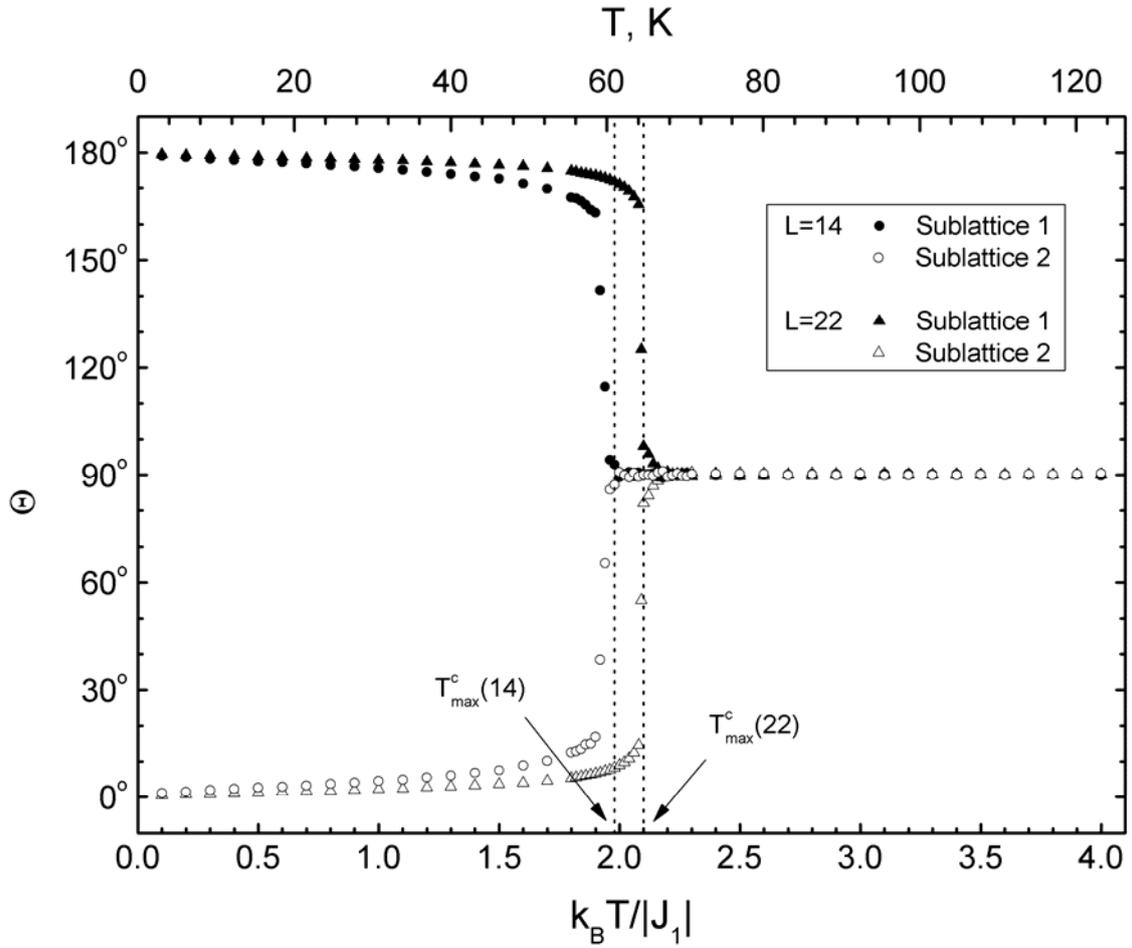

**Fig. 5.** Temperature dependence of the angle θ between the direction of the sublattice magnetization vectors and the $z$ axis for particles with the linear sizes L = 14 (the filled and open circles are for sublattices 1 and 2, respectively) and 22 (the filled and open triangles are for sublattices 1 and 2, respectively).

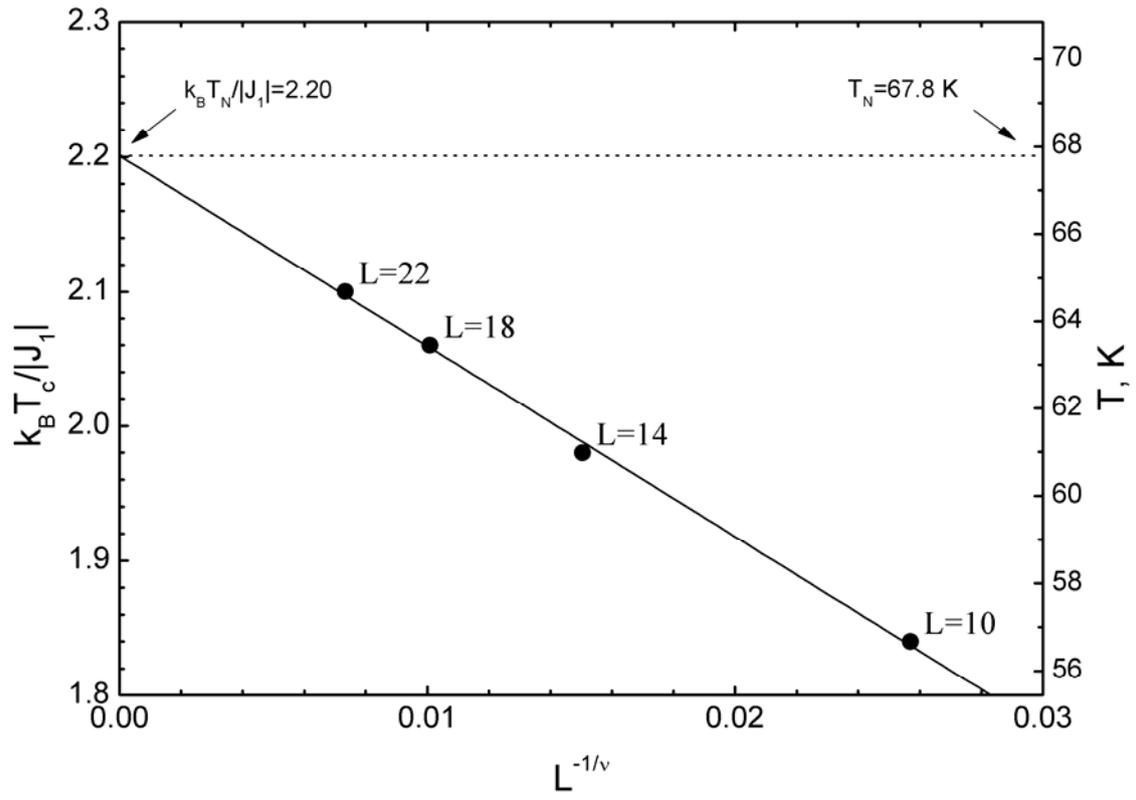

**Fig. 6**. Critical temperature versus linear particle sizes.